# MINIMIZATION OF TRANSVERSE WAKEFIELDS IN THE NLC ACCELERATOR STRUCTURES*

R.M. Jones, R.H. Miller, and J.W. Wang,
Stanford Linear Accelerator Center, Stanford, CA 94309

## Abstract

The progress of a multiple bunches of electrons through several thousand accelerator structures results in a wakefield which if left unchecked will kick successive bunches off the axis of the accelerator and can at the very least dilute the final luminosity of the final colliding beams, or at worst can lead to a BBU (Beam Break Up) instability. In order to damp the wakefields to acceptable levels for travelling wave structures we detune the frequencies of the cells and we couple out the field to four adjacent manifolds. Optimizing the manifold-cell coupling for several hundred cells and changing the bandwidth parameters of the distribution has in previous structures been achieved by a process of trial and error. Here, we report on an optimized Fortran code that has been specifically written with the aim minimizing the sum of the squares of the RMS and standard deviation of the sum wakefield. Sparse matrix techniques are employed to reduce the computational time required for each frequency step. The wakefield is minimized whilst ensuring that no significant local surface heating occurs due to slots cuts into the accelerator cells to couple out the wakefield.

Paper presented XXI International Linear Accelerator Conference (LINAC2002),
Hotel Hyundai in Gyeongju, Korea.
August 19 – August 23rd, 2002

*This work is supported by DOE grant number DE-AC03-76SF00515

# MINIMIZATION OF TRANSVERSE WAKEFIELDS IN THE NLC ACCELERATOR STRUCTURES[†]

R.M. Jones, R.H. Miller and J.W. Wang; SLAC, ARDA, Menlo Park, USA

Abstract

The progress of a multiple bunches of electrons through several thousand accelerator structures results in a wakefield which if left unchecked will kick successive bunches off the axis of the accelerator and can at the very least dilute the final luminosity of the final colliding beams, or at worst can lead to a BBU (Beam Break Up) instability. In order to damp the wakefields to acceptable levels for travelling wave structures we detune the frequencies of the cells and we couple out the field to four adjacent manifolds. Optimizing the manifold-cell coupling for several hundred cells and changing the bandwidth parameters of the distribution has in previous structures been achieved by a process of trial and error. Here, we report on an optimized Fortran code that has been specifically written with the aim minimizing the sum of the squares of the RMS and standard deviation of the sum wakefield. Sparse matrix techniques are employed to reduce the computational time required for each frequency step. The wakefield is minimized whilst ensuring that no significant local surface heating occurs due to slots cuts into the accelerator cells to couple out the wakefield.

## 1. INTRODUCTION

The main X-band linacs in the NLC have been redesigned from 1.8 meters structures operating in the $2\pi/3$ accelerating mode to 60cm structures operating in the $5\pi/6$ accelerating mode. A new structure was required as electrical breakdown was found to occur in RDDS1 [1]. The shorter high phase advance structure under consideration in this paper is known as HDDS (High Phase Advance Damped Detuned Structure) and it consists of 55 cells in which the average group velocity is 2% of the velocity of light. This structure incorporates a Sech distribution of frequencies and manifold damping is achieved by a series of slots cut into the accelerator to couple out the wakefield. Recent experiments are encouraging as higher gradients with reduced breakdown rates and minimal damage have been obtained with short low group velocity structures. [2].

In all previous designs the electromagnetic field is enhanced in the direct vicinity of the slots and this gives rise to an increased pulse temperature rise. In order to reduce the temperature rise the coupling slots have been redesigned so that they are now "pie" shaped (as they taper in towards the cell). This results in a more uniform distribution of fields along the slots. The temperature rise is calculated to be 25° C (reduced from ~70° C).

The new dipole wakefield is calculated using the method outlined in [3] for the first band for three separate cases: a single structure in which all 55 cells are coupled to the manifold, a structure in which 2 cells are decoupled from either end of the accelerator and, a series of 3 structures with 3-fold interleaving of cell frequencies (each of which has two cells decoupled). In all cases the wakefield is optimized and it is described in the following section. The beam dynamics resulting from the three separate cases are discussed in the final section.

## 2. WAKEFIELD OPTIMIZATION

The group velocity of the fundamental mode imposes a restriction on the bandwidth of the dipole mode. In the present design the bandwidth is limited to approximately 10% or less of the central frequency. The new structure has 55 cells compared to 206 in the original RDDS1 [4] and as there are less modes in a similar bandwidth then the modes do not overlap as effectively. Thus, to achieve similar wakefields as RDDS1 requires larger damping than was previously used and this necessitates larger cell-to-manifold coupling and a careful optimization of the frequency distribution.

In the design of all structures we minimize the following "cost" function:

$$\Delta = W_1^2 + S_{RMS}^2 + S_\sigma^2 \qquad (2.1)$$

Here, the first term $W_1$ is the wakefield at the first trailing bunch due to the first accelerated bunch and from a "daisy-chain" model [5] it is found to be a major indicator as to whether BBU occurs. The remaining terms are dependent on the sum wakefield, which at a particular bunch location, is the sum of the wakefield of all preceding bunches. The RMS of the sum wakefield, $S_{RMS}$, has been found to be an indicator as to whether or not BBU will occur and in practice it has been found that provided its value is less than unity then BBU is unlikely to occur. The final quantity is $S_\sigma$ the standard deviation of the sum wakefield and this is an indicator of alignment tolerances [6].

In order to minimize $\Delta$, the bandwidth and $\sigma$ of the frequency distribution of cells are varied. We chose an initial distribution of the form $(\text{Sech}(f/\sigma))^{1.5}$ and this distribution has similar properties to a Gaussian but with a slightly improved wakefield. The results of this optimization process are shown in Figs. 1, in which three simulations are illustrated. Provided all cells are coupled to the manifold then a well-damped wakefield is obtained.

[†] Supported by DOE grant number DE-AC03-76SF00515

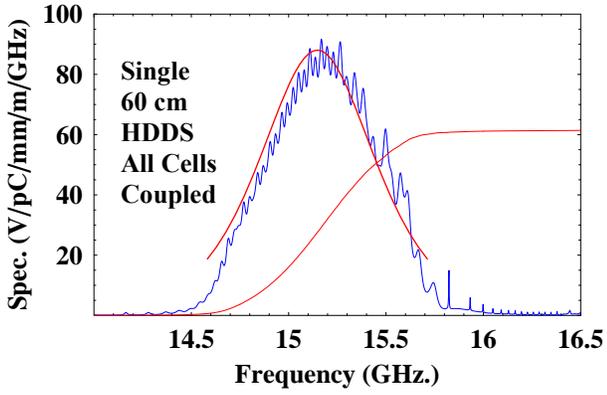
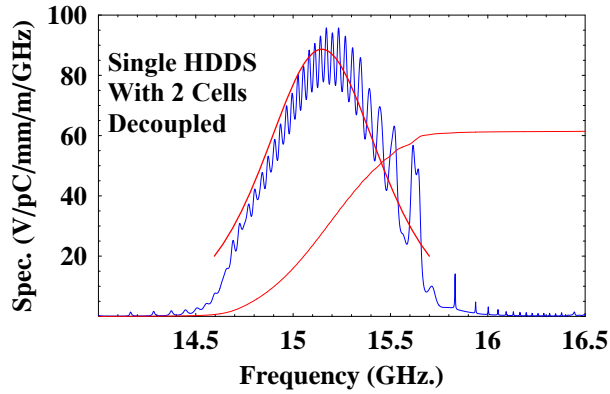
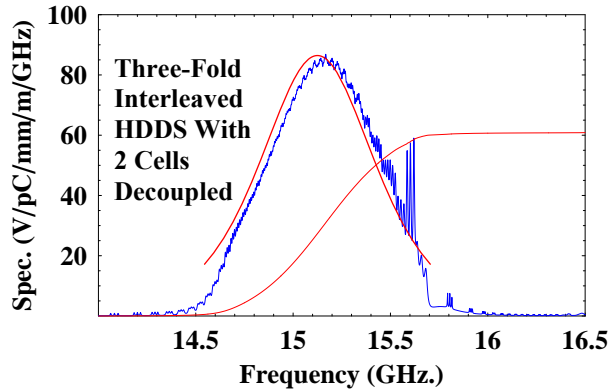
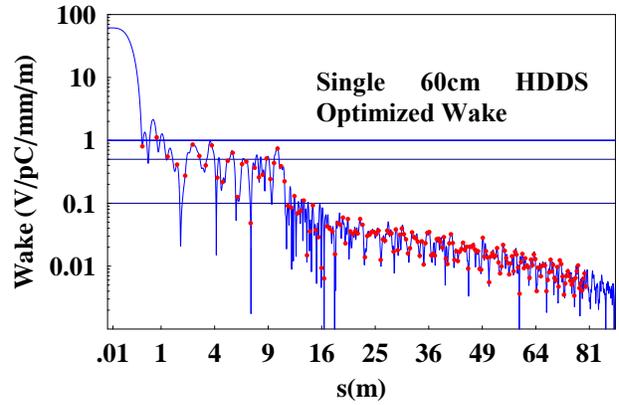
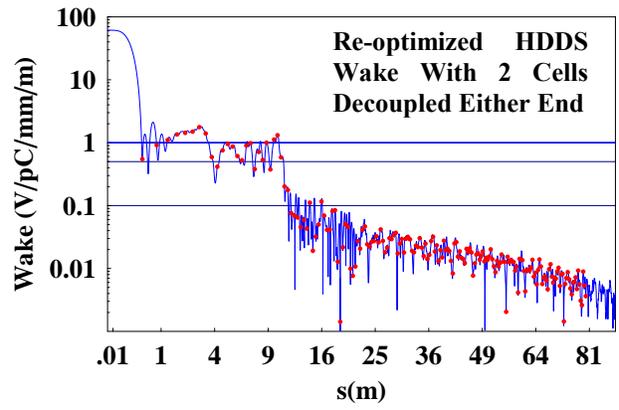
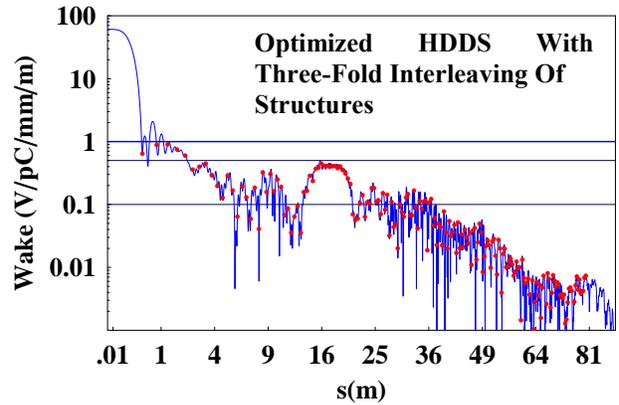

Figure 1. Spectral function for HDDS under three different conditions. Shown uppermost is a single structure in which the cost function has been minimized when all the cells are coupled to the manifold. The bandwidth is 7.46% and the total frequency width is 3.38$\sigma$. Decoupling 2 cells from the manifold at either end of the structure is shown in the middle figure and it clearly degrades the spectral function significantly. The re-optimised bandwidth is 7.29% and 3.30 units of $\sigma$. The optimized spectral function with 2 cells decoupled and 3 fold interleaving is shown lowermost for an initial bandwidth 7.67% and 3.48$\sigma$

Figure 2. Envelope of wake function for the conditions given in the adjacent figure. The points indicate the location of the bunches. Each bunch is separated from its neighbour by 42cm and there are 192 of them in each bunch train. The wake incorporating 3-fold interleaving of structures is below 1V/pC/mm/m for all bunches. Beam dynamics simulations have indicated that provided the total (including higher bands) wake is lower than unity then BBU will be prevented. The three-fold interleaved case, shown lowermost required a 50% reduction in cell-to-manifold coupling compared to the non-interleaved case. This allows for some flexibility in the design of the coupling slots.

The impact on the wakefield of decoupling two cells on either cells of the structure is shown in the centre of Fig. 2 and it is seen to degrade the wakefield significantly. In earlier structures we were required to decouple cells from the manifold in order to facilitate attachment of fundamental and higher order mode couplers. For HDDS1 a redesigned fundamental mode coupler [7], with reduced fields will be used, and it may be possible to couple all 55 cells. However, if 2 or more cells are forced to be decoupled then three-fold interleaving of structures will be required and the wakefield for this situation is shown lowermost in Fig. 2.

## 3. BEAM DYNAMICS

An indication as to whether BBU occurs is provided by $S_{RMS}$ and this is illustrated in Fig. 3 as a function of a small fractional error in the bunch spacing (equivalent to a systematic error in the frequencies of all cells). In all three designs the minimum value of $S_{RMS}$ is seen to lie at design value of the bunch spacing. The single partially coupled structure is very sensitive to systematic errors as $S_{RMS}$ rises to unity and above with frequency errors as little as 3 MHz. Both the fully-coupled and the partially coupled with three-fold interleaving of structures can tolerate frequency errors in excess of 15 MHz before the BBU instability is likely to occur.

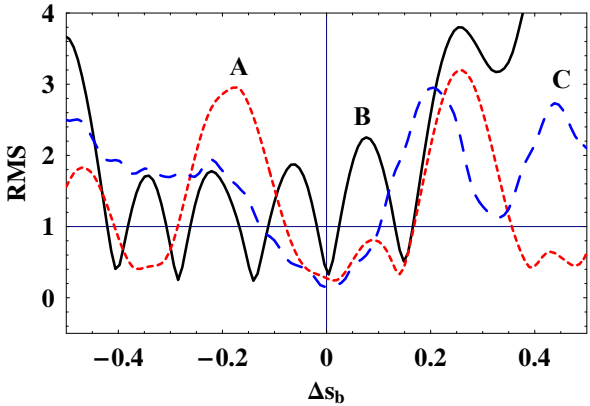

Figure 3 $S_{RMS}$ as a function of the fractional change in the bunch spacing for three cases. The red solid line (A) is obtained for all cells coupled to the damping manifold of the structure, black short dashes (B) to a single structure in which two cells are decoupled from the manifold at either end of the structure, long blue dashed (C) to 2 cells decoupled and three-fold interleaving of structures.

Further evidence as to whether BBU is occurring is provided by tracking the normalized emittance with respect to the centroid of the beam down the complete linac. The result of tracking the progress of a beam initially offset by 1μm using the computer code LIAR [8] is shown in Fig. 4. It is clear that both the single fully coupled and three-fold interleaved structure gives rise to little emittance growth. Indeed, the final emittance dilution results almost entirely from the short range transverse wakefield which is also included in the simulation.. However, the single structure with cells decoupled clearly gives rise to unacceptable emittance growth as BBU is seen to be occurring (phase space curves, not shown, also confirm this).

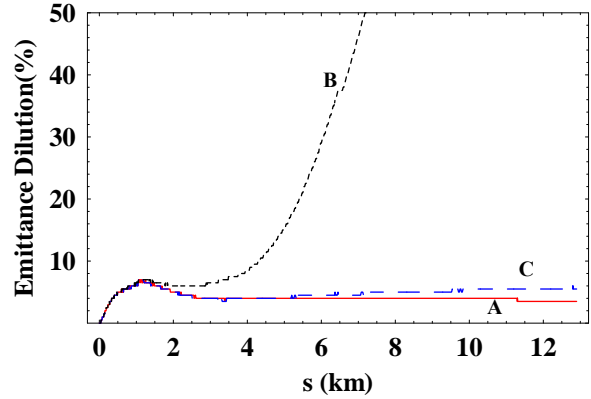

Figure 4. Simulation of $\Delta\varepsilon/\varepsilon_0$, the percentage emittance growth for the structures given in Fig. 3 with $\Delta s_b = 0$.

In order to assess the worst case scenario systematic errors were assigned according to the nearest peaks in $S_{RMS}$. The frequency errors assigned in cases A, B and C are 39, 11.4 and 30.7 MHz respectively. Even though these large errors are unlikely to occur in practice, Fig 5 shows interestingly, that the fully-coupled single structure only degrades the beam emittance by an additional 15%.

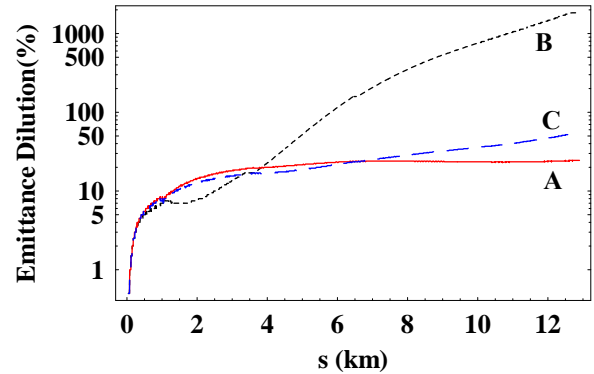

Figure 5. $\Delta\varepsilon/\varepsilon_0$ for similar conditions to those given in Fig 4 except that $S_{RMS}$ has been maximized for each case.

## 5. REFERENCES

[1] C. Adolphsen et al, PAC2001, SLAC-PUB-8901
[2] V. Dolgashev et al, TH466, these proceedings
[3] R.M. Jones et al, LINAC96, SLAC-PUB-7287
[4] J.W. Wang et al, LINAC2000, SLAC-PUB-8583
[5] K. Thompson and R.D. Ruth, 1989, SLAC-PUB-4801
[6] R.M. Jones et al, PAC99, SLAC-PUB-8101
[7] C. Nantista, private communication .
[8] R. Assman et al, LIAR, SLAC-PUB AP-103